\documentclass{article}

\usepackage{axodraw}
\usepackage{color}
\def\la{\langle}
\def\ra{\rangle}

\def\beq{\begin{equation}}
\def\eeq{\end{equation}}
\def\bea{\begin{eqnarray}}
\def\eea{\end{eqnarray}}
\def\barr{\begin{array}}
\def\earr{\end{array}}

\def\op{{\mathcal{O}}}

\def\ChPT{$\chi PT$}

\begin{document}
%%%%%%%%%%% Titlepage

\begin{titlepage}
\begin{flushright}
INT-PUB 03-20\\
NT@UW-03-029\\
SISSA 99/03 EP\\
UW/PT 03-24\\
\end{flushright}
\vskip 0.5cm
\begin{center}
%\vskip0.2cm the $\Delta I=1/2$ case}
{\Large \bf A note on the power divergence in lattice calculations
of\\ \vspace{0.2cm}
$\Delta I = 1/2$ $K\rightarrow\pi\pi$ amplitudes at $M_{K}=M_{\pi}$
%\\ \vspace{0.3cm}
} \vskip1cm {\large\bf Maarten~Golterman$^a$,
C.-J.David~Lin$^{b,c}$, Elisabetta~Pallante$^d$}\\ \vspace{.5cm}
{\normalsize {\sl 
$^a$ Department of Physics and Astronomy, San Francisco State University,\\
1600 Holloway Ave, San Francisco, CA 94132, USA.\\ 
\vspace{.2cm}
$^b$ Department of Physics,  
University of Washington, Seattle, WA 98195-1550, USA.\\ 
\vspace{.2cm}
$^c$ Institute for Nuclear Theory, 
University of Washington, Seattle, WA 98195-1550, USA.\\ 
\vspace{.2cm}
$^d$ SISSA and INFN, Sezione di Trieste, Via Beirut 2-4,
34013, Trieste, Italy.
}}

\vskip1.0cm {\large\bf Abstract:\\[10pt]} \parbox[t]{\textwidth}{{
In this note, we clarify a point concerning the power divergence
in lattice calculations of 
$\Delta I = 1/2$ $K\rightarrow\pi\pi$ decay amplitudes.  
There have been worries that this divergence
might show up in the Minkowski amplitudes at 
$M_{K}=M_{\pi}$ with all the mesons at rest.
Here we demonstrate, via an explicit calculation in leading-order
Chiral Perturbation Theory, that the power divergence is
absent at the above kinematic point, as predicted by CPS symmetry.
}}
\end{center}
\end{titlepage}

%%%%%%%%%%%%%%

The subtraction of a power divergence, which arises via the mixing
of dimension-six four-fermion operators with those of lower 
dimension 
has been one of the central issues in lattice calculations of 
$\Delta I = 1/2$ $K\rightarrow\pi\pi$ amplitudes.  This power
divergence is of course unphysical, and can
be related to a shift of the vacuum due to the
inclusion of the weak interaction in Chiral Perturbation Theory ($\chi$PT)
\cite{Bernard:wf,Crewther:1985zt, Leurer:1987ih, Kambor:1989tz}.  
It results in the so-called tadpole
operators, which contribute to the processes $K^{0}\rightarrow |0\ra$
and $\bar{K}^{0}\rightarrow |0\ra$, in $\chi$PT with weak interactions.

%\medskip

As argued in Ref. \cite{Bernard:1987pr}, this power divergence should be
absent for $K\rightarrow\pi\pi$ amplitudes when $m_{s}=m_{d}=m_{u}$
($m_{u,d,s}$ are the masses of $u$, $d$ and $s$ quarks), 
due to the exact CPS symmetry \cite{Bernard:wf} of the four-fermion
operators that mediate $K\rightarrow\pi\pi$ decays.  
In Ref. \cite{Golterman:1999hv}, it was argued
that the power divergence indeed does disappear 
in {\sl Euclidean space} at $M_{K}=M_{\pi}$.
However, a naive calculation in Minkowski space suggests that this
power divergence might still be present at $M_K=M_\pi$ when all mesons 
are at rest.  The issue is relevant, as it has been proposed that this 
unphysical kinematic point can be used to extract the low-energy constants 
relevant for $\Delta I=1/2$ $K\to \pi\pi$ to order $p^4$ in $\chi$PT 
\cite{Laiho:2003uy,Laiho:2002jq}.\footnote{It follows from our analysis 
that the
low-energy constants $\alpha_2$ and $e^r_{1,2,5}$ should not appear
in Eq.~(31) of Ref. \cite{Laiho:2002jq}.}

%\medskip

In this note, we show, via an explicit calculation in $\chi$PT, 
that also in {\sl Minkowski} space the power divergence is not
present in $\Delta I=1/2$ $K\to\pi\pi$ amplitudes
at $M_{K}=M_{\pi}$, with all mesons at rest.
Since it has already been argued in Ref. \cite{Lin:2003tn} that
the $\Delta I = 1/2$ 
$K\rightarrow\pi\pi$ amplitudes in partially quenched $\chi$PT 
at the kinematic point $M_{K}=M_{\pi}$ suffer from problems
related to the lack of unitarity 
\cite{Golterman:1999hv, Lin:2002aj, Bernard:1995ez},
we concentrate here on full QCD. Our conclusions on the 
power divergence will however not change in the (partially) quenched case.

To simplify the discussion, we only
consider weak operators in the $(8,1)$ irrep of 
$SU(3)_{L}\times SU(3)_{R}$.  The weak mass operator
in this irrep at $\op(p^{2})$ in the chiral expansion is
\beq
\label{eq:tadpole81}
 \op^{(8,1)}_{2} = \alpha_{2}\times \left \{
  2 B_{0} {\mathrm{Tr}}\left [ \lambda_{6}
  \left ( {\mathcal{M}}^{\dagger} \Sigma
   + \Sigma^{\dagger} {\mathcal{M}}\right )\right]
 \right \} ,
\eeq
where $\alpha_{2}$ is the (power-divergent)
low-energy constant associated with this 
operator, $B_{0} = -\la 0|\bar{u}u+\bar{d}d|0\ra/f^{2}$
(in the chiral limit), $\lambda_{6}$ is
a Gell-Mann matrix, ${\mathcal{M}}$ is the quark-mass matrix and 
$\Sigma$ is the standard non-linear Goldstone field.  

%\medskip

We first observe that CPS symmetry
implies that the parity-odd part of this
operator is proportional to $m_s-m_d$.  
In fact,
\beq
\label{eq:tadpole81_2}
 \op^{(8,1)}_{2} = \alpha_{2}\times \left \{
   B_{0} (m_s+m_d){\mathrm{Tr}}\left [ \lambda_{6}\left ( \Sigma
   + \Sigma^{\dagger}\right ) \right]
+iB_{0} (m_s-m_d){\mathrm{Tr}}\left [ \lambda_{7}\left ( \Sigma
   - \Sigma^{\dagger}\right ) \right]
 \right \} .
\eeq
Therefore, at $m_s=m_d$ the parity-odd part of the {\sl operator} 
vanishes, and thus its $K\to\pi\pi$ matrix 
element should vanish as well for $M_K=M_\pi$.  This was confirmed by an
explicit calculation in Euclidean space (as reported in Ref.
\cite{Golterman:1999hv}), and should
be true in Minkowski space as well. 

%%%

%%
\begin{center}
\begin{figure}
\begin{center}
\begin{picture}(400,250)(0,-65)
\ArrowLine(75,30)(135,60) \ArrowLine(75,30)(135,0)
\ArrowLine(0,30)(75,30)
\GCirc(75,30){5}{0.8} \GCirc(135,60){2}{0.1}\GCirc(135,0){2}{0.1}
\GCirc(0,30){2}{0.1}
\Text(35,22)[t]{$K^{0}$}
\Text(105,60)[t]{$\pi^{+}$}
\Text(105,10)[t]{$\pi^{-}$}
\Text(75,-20)[t]{(a)}
\ArrowLine(325,30)(385,60) \ArrowLine(325,30)(385,0)
\ArrowLine(250,30)(325,30) \DashArrowLine(270,60)(325,30){3}
\GCirc(385,60){2}{0.1}\GCirc(385,0){2}{0.1}
\GCirc(250,30){2}{0.1}\GCirc(270,60){5}{0.8}
\GBox(321,26)(329,34){0.1}
\Text(285,22)[t]{$K^{0}$}
\Text(305,60)[t]{$\bar{K}^{0}$}
\Text(355,60)[t]{$\pi^{+}$}
\Text(355,10)[t]{$\pi^{-}$}
\Text(325,-20)[t]{(b)}
\end{picture}
\end{center}
\caption{\label{fig:diagrams}Diagrams involving the weak mass operator
at the lowest order in the 
chiral expansion for the $\Delta I = 1/2$ $K\rightarrow\pi\pi$
amplitudes.  The gray circles represent the operator
$\op^{(8,1)}_{2}$, and the square is the
$K^{0}\bar{K}^{0}\rightarrow\pi^{+}\pi^{-}$ vertex from the lowest-order
strong chiral Lagrangian.  The dashed line in (b) indicates that the 
$\bar{K}^{0}$ could
be off-shell, while all the other mesons are always on-shell.}
\end{figure}
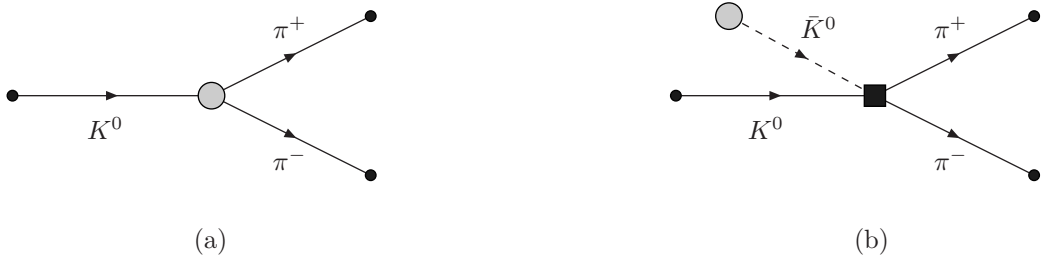
\end{center}
%%
%%%

At leading order in the chiral expansion,
$\op^{(8,1)}_{2}$ contributes to the $K\rightarrow\pi\pi$ amplitudes
via the diagrams in Fig. \ref{fig:diagrams}, where the gray 
circles represent the weak mass operator, and the square 
is from the leading-order strong chiral Lagrangian.  In diagram (b), 
there is a pole associated with the $\bar{K}^{0}$ propagator 
(the dashed line in Fig. \ref{fig:diagrams}), which takes the form
\beq
\label{eq:Kbar_prop}
 \frac{i}{(M_{K}-2 M_{\pi})^{2} - M_{K}^{2} + i \epsilon},
\eeq
when all the other three on-shell particles are at rest.  For fixed $M_K\ne 
M_\pi$, one may take $\epsilon\to 0$ at any stage of the calculation, since
the denominator of Eq.~(\ref{eq:Kbar_prop}) does not vanish in that case.
However, for $M_K=M_\pi$, the $i\epsilon$ prescription is needed in order
to define the propagator, and should be taken to zero 
{\sl only at the end of the calculation}.  In that case, one finds:
\beq
\label{eq:correct_o2Kpipi}
 \la\pi^{+}\pi^{-}|\op^{(8,1)}_{2}|K^{0}\ra_{M_{K}=M_{\pi}}
  = \lim_{\epsilon\rightarrow 0}\left \{ \lim_{M_{K}\rightarrow M_{\pi}}
  \alpha_{2}\times \left ( M^{2}_{K}-M^{2}_{\pi}\right )\left[
  \frac{8\left ( 
  3 M_{\pi} (M_{K} - 2 M_{\pi}) - i\epsilon
  \right )}{3 f^{3} (4 i M_{\pi} (M_{K} - M_{\pi}) + \epsilon)}
  \right ]\right \}= 0 ,
\eeq
which indicates that there is no need to perform the subtraction of a 
power divergence.

\medskip

Let us discuss this claim in some more detail.  We begin by noting that
it was shown long ago \cite{Bernard:wf, Crewther:1985zt, Leurer:1987ih}
that if the weak mass term $\int d^4x{\cal O}^{(8,1)}_2$ is treated as
a perturbation to the strong chiral Lagrangian, 
it does not have any observable
effect.  However, here we consider the unphysical situation of
an energy non-conserving matrix
element of ${\cal O}^{(8,1)}_2$ (corresponding to the insertion of this
operator at a fixed time, see below), and the above consideration does not
apply.

The factor $(M^{2}_{K}-M^{2}_{\pi})$
on the right-hand side of Eq. (\ref{eq:correct_o2Kpipi}) 
originates from the CPS symmetry of the operator
({\it c.f.} Eq. (\ref{eq:tadpole81_2})), while the quantity in the 
square brackets is determined by the kinematics.  This latter quantity
indeed diverges in the limit $M_{K}\rightarrow M_{\pi}$ (and $\epsilon\to 0$).
That this is exactly what one expects to happen follows because
the $\bar{K}^{0}$ propagator in Fig. \ref{fig:diagrams}(b) 
goes on-shell without being amputated.  In fact, 
Fig. \ref{fig:diagrams}(b) also represents the process of $K^0$-$\bar{K}^{0}$
scattering into $\pi^+$-$\pi^-$, but in that case in order to
obtain a finite amplitude, the LSZ reduction formula tells us to amputate
the $\bar{K}^{0}$ external
leg, before putting it on-shell.  Since in our case this leg is not
amputated, the diagram is divergent in the on-shell limit.
In the case that $K^{0}$, $\pi^{+}$ and $\pi^{-}$ are
all at rest, this ``$\bar{K}^{0}$ on-shell'' point coincides with the
limit $M_{K}\rightarrow M_{\pi}$, and CPS symmetry prevents the
divergence from happening:  the amplitude actually vanishes at
$M_{K}=M_{\pi}$.  

However, one may consider the following more general situation.
Consider for instance 
kinematics with $K^{0}$ at rest but $\pi^{+}$ and $\pi^{-}$
carrying spatial momenta $\vec{p}$ and $-\vec{p}$ respectively. 
In that case, the ``$\bar{K}^{0}$ on-shell'' point is at 
$M_{K}=E_{\pi}=\sqrt{M_{\pi}^{2}+|\vec{p}|^{2}}$, and diagram
\ref{fig:diagrams}(b) is proportional to $\vec{p}^{2}/\epsilon$
at this point.  The extra factor $(M^{2}_{K}-M^{2}_{\pi})$ clearly
does not help in this case, and the 
divergence occurs  of course for exactly the same reason as described
above.

We gain more insight by considering the amplitude in position space,
as in Fig. \ref{fig:correlators}(b).  This diagram contains a factor
${\mathrm{e}}^{-i M_K |t_w-t_s|}$ from the $\bar{K}^{0}$ propagator, 
where $t_w$ is the location (in time)
of the weak operator ${\cal O}^{(8,1)}_2$ (taken as $t_w=0$ in the
diagram), and $t_s$ is the location of the strong vertex.  The LSZ
prescription for this $\bar{K}^{0}$ line corresponds to taking a
Fourier transform with respect to $t_w$, and putting the corresponding
momentum on-shell.  For this to work, the integral over $t_w$ needs
to be regulated by replacing $M_K\to M_K-i\epsilon$, and this is 
precisely what leads to the $i\epsilon$ prescription 
in Eq. (\ref{eq:Kbar_prop}).  It follows that the divergence
encountered here is regulated by considering the amplitude at finite
$t_w$ (by time-translation invariance we may choose $t_w=0$).  This
is of course what one does anyway in a lattice computation of this
amplitude.  It is therefore instructive to consider this amplitude
in position space rather than momentum space \cite{Golterman:1999hv}, 
which is what we will do next.  
\begin{center}
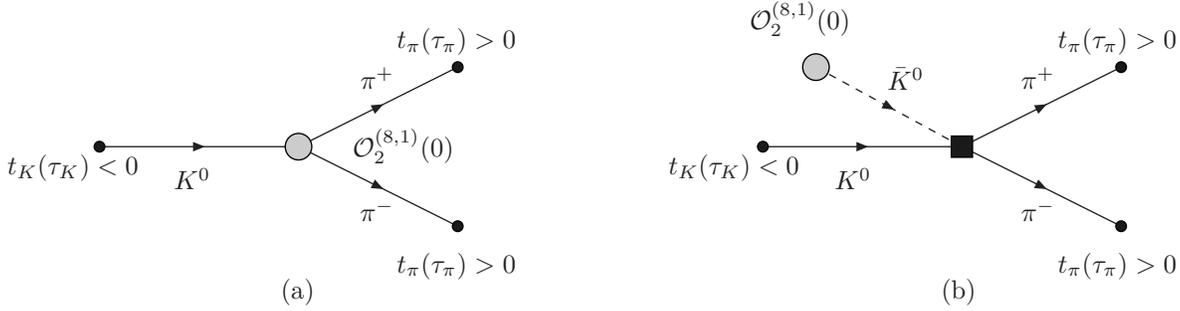
\begin{figure}
\begin{center}
\begin{picture}(400,250)(0,-65)
\ArrowLine(75,30)(135,60) \ArrowLine(75,30)(135,0)
\ArrowLine(0,30)(75,30)
\GCirc(75,30){5}{0.8} \GCirc(135,60){2}{0.1}\GCirc(135,0){2}{0.1}
\GCirc(0,30){2}{0.1}
\Text(35,22)[t]{$K^{0}$}
\Text(105,60)[t]{$\pi^{+}$}
\Text(105,10)[t]{$\pi^{-}$}
\Text(75,-20)[t]{(a)}
\Text(115,37)[t]{$\op^{(8,1)}_{2}(0)$}
\Text(-10,27)[t]{$t_{K} (\tau_{K}) < 0$}
\Text(135,75)[t]{$t_{\pi} (\tau_{\pi}) > 0$}
\Text(135,-10)[t]{$t_{\pi} (\tau_{\pi}) > 0$}
\ArrowLine(325,30)(385,60) \ArrowLine(325,30)(385,0)
\ArrowLine(250,30)(325,30) \DashArrowLine(270,60)(325,30){3}
\GCirc(385,60){2}{0.1}\GCirc(385,0){2}{0.1}
\GCirc(250,30){2}{0.1}\GCirc(270,60){5}{0.8}
\GBox(321,26)(329,34){0.1}
\Text(285,22)[t]{$K^{0}$}
\Text(305,60)[t]{$\bar{K}^{0}$}
\Text(355,60)[t]{$\pi^{+}$}
\Text(355,10)[t]{$\pi^{-}$}
\Text(325,-20)[t]{(b)}
\Text(265,85)[t]{$\op^{(8,1)}_{2}(0)$}
\Text(240,27)[t]{$t_{K} (\tau_{K}) < 0$}
\Text(385,75)[t]{$t_{\pi} (\tau_{\pi}) > 0$}
\Text(385,-10)[t]{$t_{\pi} (\tau_{\pi}) > 0$}
\end{picture}
\end{center}
\caption{\label{fig:correlators}Diagrams involving the weak mass operator
at the lowest order in the chiral expansion for the correlator
$\la 0 |\pi^{+}\pi^{-} Q^{(8,1)} \bar{K}^{0}|0\ra$.
The gray circles represent the weak mass operator
$\op^{(8,1)}_{2}$, and the square is the
$K^{0}\bar{K}^{0}\rightarrow\pi^{+}\pi^{-}$ vertex from the lowest-order
strong chiral Lagrangian.  The dashed line in (b) means $\bar{K}^{0}$ could
be off-shell, while all the other mesons are always on-shell.
The weak operator
is at the space-time origin.  $K^{0}$ is created at $t_{K}$ ($\tau_{K}$),
and the pions are annihilated at $t_{\pi}$ ($\tau_{\pi}$) in Minkowski
(Euclidean) space.}
\end{figure}
\end{center}

\medskip

Since we take all our mesons to have vanishing spatial momentum,
we will consider
the relevant correlators in the time-momentum
representation, {\it i.e.} study the correlators as functions of
3-momentum and time.  In this setup, a free meson propagator with energy
$E_{\vec{p}}=\sqrt{m^{2}+|\vec{p}|^{2}}$ ($m$ is the mass and $\vec{p}$ is the
3-momentum of the meson) is
\bea
 \frac{{\mathrm{e}}^{-i E_{\vec{p}} |t|}}{2 E_{\vec{p}}} , 
 & & ({\mathrm{Minkowski}})\nonumber
 \\ & &\nonumber \\
 \frac{{\mathrm{e}}^{- E_{\vec{p}} | \tau |}}{2 E_{\vec{p}}}
, & & ({\mathrm{Euclidean}})\nonumber
\eea
where $t$ ($\tau$) is the Minkowski (Euclidean) time.  The time dependence 
of the Minkowski expression is of course in accordance with the $i\epsilon$
prescription of Eq.~(\ref{eq:Kbar_prop}). 
We now consider the Minkowski correlator,
\bea
 C_{2} &=& \la 0 | \pi^{+}_{\vec{0}}(t_{\pi})\pi^{-}_{\vec{0}}(t_{\pi}) 
  \op^{(8,1)}_{2}(0) K^{0}_{\vec{0}}(t_{K})|0\ra ,
\eea
and its Euclidean counterpart,
\bea
 {\mathcal{C}}_{2} &=& \la 0 | \pi^{+}_{\vec{0}}(\tau_{\pi})
   \pi^{-}_{\vec{0}}(\tau_{\pi}) 
  \op^{(8,1)}_{2}(0) K^{0}_{\vec{0}}(\tau_{K})|0\ra .
\eea
For simplicity, we choose to annihilate the two pions at the same time,
and assume that $t_{K} (\tau_{K}) < 0$ and $t_{\pi} (\tau_{\pi}) > 0$. 
The weak operator is inserted at time $t_w=0$ ($\tau_w=0$).
All particles are at rest (as indicated by the subscripts $\vec{0}$).
The relevant
diagrams for the above correlators are shown in Fig. \ref{fig:correlators}.
In the following, we only present the result in Minkowski space, but stress
that the calculation in Euclidean space is virtually identical, and 
leads to the same conclusion \cite{Golterman:1999hv}.

The contribution from diagram \ref{fig:correlators}(a) to the correlator
$C_{2}$ is:
\bea
\label{eq:o812_a}
 C_{2(a)} &=& \frac{-8 i\alpha_{2}}{3 f^{3}}
 \left ( M^{2}_{K}-M^{2}_{\pi}\right )
 \left [\frac{{\mathrm{e}}^{-iM_{K}|t_{K}|}{\mathrm{e}}^{-2iM_{\pi}t_{\pi}}}
      {(2M_{K})(2M_{\pi})(2M_{\pi})}\right ] ,
\eea
while diagram \ref{fig:correlators}(b) leads to
\bea
\label{eq:o812bar_b}
 C_{2(b)} &=& \frac{4 i\alpha_{2}}{3 f^{3}}
 \left ( M^{2}_{K}-M^{2}_{\pi}\right ) 
\frac{i}{(2M_{K})(2M_{K})(2M_{\pi})(2M_{\pi})}\nonumber\\
 & &\times\int d t_{s} \mbox{ }
 {\mathrm{e}}^{-iM_{K} |t_{s} - t_{K}|} {\mathrm{e}}^{-iM_{K} |t_{s}|}
  {\mathrm{e}}^{-2iM_{\pi} |t_{\pi} - t_{s}|}\nonumber\\
 & &\mbox{ } \times \left \{
  M^{2}_{K} \left [ 1 + \epsilon(t_{s})\epsilon(t_{s}-t_{K})\right ] 
 + M_{K} M_{\pi} \left [ \epsilon(t_{s}) + \epsilon(t_{s} - t_{K})
  \right ] \epsilon(t_{\pi}-t_{s}) + 2 M^{2}_{\pi}  
 \right \},
\eea
where $t_{s}$ is the time-component of the space-time position of
the strong chiral Lagrangian vertex in this diagram.  The function
$\epsilon (t)$ is defined as:
\beq
 \epsilon (t) = \Bigg \{ \begin{array}{r} +1, \mbox{ } t > 0 .\\ 
  \mbox{ }\\
      -1,\mbox{ }t < 0 .\end{array} 
\eeq
In the above two equations, only the integral of $t_{s}$ between 0 
and $t_{\pi}$ can result in a ``vanishing denominator'' when 
$M_{K} \rightarrow M_{\pi}$.  Explicitly, it is:
\bea
 \label{eq:o812bar_0_tpi}
 C_{2(b)}|_{0\rightarrow t_{\pi}} &=& -i\, C_{2(a)}\times
 \left (\frac{ M^{2}_{K} +  M^{2}_{\pi} + M_{K} M_{\pi}}{2M_{K}}
 \right )\nonumber\\
 & &\mbox{ }\mbox{ }\mbox{ }\mbox{ }\mbox{ }\mbox{ }
    \mbox{ }\mbox{ }\mbox{ }
   \times\left\{ \frac{1}{-2i(M_{K}-M_{\pi})}\left [
  {\mathrm{e}}^{-2i(M_{K}-M_{\pi})t_{\pi}} - 1
 \right ]\right\} .
\eea
When $M_{K}\rightarrow M_{\pi}$, the factor
\bea
 & &\frac{1}{-2i(M_{K}-M_{\pi})}\left [
  {\mathrm{e}}^{-2i(M_{K}-M_{\pi})t_{\pi}} - 1
 \right ]\nonumber
\eea
is just $t_{\pi}$.  Therefore, for finite $t_{\pi}$ (or finite $\tau_{\pi}$
in Euclidean space), $C_{2}$ (or ${\mathcal{C}}_{2}$ in Euclidean space) 
vanishes at $M_{K}=M_{\pi}$ (with both $C_{2(a)}=0$ and $C_{2(b)}=0$ 
separately) due to the explicit factor of $M^{2}_{K}-M^{2}_{\pi}$, 
and there is no power 
divergence.  This conclusion remains true to all orders in $\chi$PT.

\medskip

To conclude, we would like to discuss in some more detail why the factor
linear in $t_\pi$ appears in Eq. (\ref{eq:o812bar_0_tpi}), even though
$C_{2(b)}$ vanishes for $M_K=M_\pi$ because of the explicit factor
$(m_s-m_d)$ in Eq. (\ref{eq:tadpole81_2}).  Omitting this factor,
our result contains a term linear in $t_\pi$ for $M_K=M_\pi$.
One would expect that if
one takes $t_{\pi}$ large after taking the limit 
$M_{K}\rightarrow M_{\pi}$, it would be necessary to unitarize $C_{2(b)}$.%
\footnote{The on-shell divergence discussed earlier has nothing to
do with this term linear in $t_\pi$, but, as explained above,
with the oscillatory behavior of ${\mathrm{e}}^{-i M_K |t_w-t_s|}$.}
Re-interpreting diagram \ref{fig:correlators}(b) 
as the lowest-order contribution in \ChPT\ (in the strong
vertex) to $K^{0}\bar{K}^{0}\rightarrow\pi^{+}\pi^{-}$ scattering
(as we did above), the term linear in $t_\pi$ can be understood
as follows.  For $M_K=M_\pi$, there is full $SU(3)$ symmetry,
and $|KK\rangle$ and $|\pi\pi\rangle$ $s$-wave,
$I=0$ states can be expressed in terms
of the ($I=0$ components of the)
irreducible states $|1\rangle$, $|8\rangle$ and $|27\rangle$
through the relations \cite{Lin:2003tn}
\bea
 |1\ra &=& \frac{1}{\sqrt{2}} |KK\ra + \frac{1}{2\sqrt{2}}|\eta\eta\ra
      + \frac{1}{2}\sqrt{\frac{3}{2}}|\pi\pi\ra , \nonumber\\
 |8\ra &=&\frac{1}{\sqrt{5}} |KK\ra + \frac{1}{\sqrt{5}}|\eta\eta\ra
      - \sqrt{\frac{3}{5}}|\pi\pi\ra , \nonumber\\
 |27\ra &=& -\sqrt\frac{3}{10} |KK\ra 
 + \sqrt{\frac{27}{40}}|\eta\eta\ra + \frac{1}{\sqrt{40}}|\pi\pi\ra , 
\eea
where 
\bea
 |KK\ra &=& \frac{1}{\sqrt{2}}\left ( |K^{0}\bar{K}^{0}\ra 
    + |K^{+}K^{-}\ra\right ) , \nonumber\\
 |\pi\pi\ra &=& \frac{1}{\sqrt{3}}\left ( |\pi^{0}\pi^{0}\ra
    + \sqrt{2}|\pi^{+}\pi^{-}\ra \right ) .
\eea
{}From these relations, it follows that
\beq
\langle\pi\pi(t=t_\pi)|KK(t=0)\rangle
=\frac{\sqrt{3}}{4}\langle 1(t=t_\pi)|1(t=0)\rangle
-\frac{\sqrt{3}}{5}\langle 8(t=t_\pi)|8(t=0)\rangle
-\frac{\sqrt{3}}{20}\langle 27(t=t_\pi)|27(t=0)\rangle .
\eeq
To leading order in \ChPT\ this expression contains
a term linear in $t_\pi$, the coefficient of which is the 
corresponding linear
combination of finite-volume two-particle energy shifts, thus
explaining how a term linear in $t_\pi$ appears in $C_{2(b)}$.
Note that the normalization of $C_{2(b)}$ has been chosen 
such that (in leading non-vanishing order)
it is independent of the spatial volume.  Higher orders
in \ChPT\ will indeed unitarize our result.

\section*{Acknowledgments}
We are grateful to Claude Bernard, Chris Dawson, Jack Laiho, Yigal Shamir,
Steve Sharpe, Amarjit Soni and Giovanni Villadoro for 
helpful discussions.  MG is supported in part by the US Department of 
Energy and CJDL acknowledges support by the US Department of 
Energy via grants DE-FG03-00ER41132, DE-FG03-96ER40956 and 
DE-FG03-97ER41014.  EP is supported in part by the Italian MURST 
under the program \textit{Fenomenologia delle Interazioni Fondamentali}.

%%%%%%%%%%%%%%%%%%%%%%%%%%%%%%%%%%%%%%%%%%%%%%%%%%%%%%%%%
%                           BIBLIOGRAPHY
%%%%%%%%%%%%%%%%%%%%%%%%%%%%%%%%%%%%%%%%%%%%%%%%%%%%%%%%%%

\end{document}